\documentclass[a4paper,11pt]{article}

\usepackage{pos}
\usepackage{multirow}
\usepackage{makecell}

\title{Leptoquark and vectorlike quark extended models as the explanation of $(g-2)_{\mu}$ anomaly}

\author*[a]{Shi-Ping He}

\affiliation[a]{Asia Pacific Center for Theoretical Physics, Pohang 37673, Korea}

\emailAdd{shiping.he@apctp.org}

\abstract{In minimal leptoquark (LQ) models, the $R_2$ and $S_1$ can be the solution to the $(g-2)_{\mu}$ anomaly because of the chiral enhancements. Here, we study the LQ and vectorlike quark (VLQ) extended models. In the one LQ and one VLQ extended models, the $(g-2)_{\mu}$ can receive the contributions from top and top partner $T$ because of the $t-T$ mixing. Besides the traditional $R_2$ and $S_1$ representations, we find that the $S_3$ LQ can also explain the anomaly when including the $(X,T,B)_{L,R}$ triplet at the same time. Moreover, we find that the LQ has the new decay channel $T\mu$ in these models.}

\FullConference{%
  41st International Conference on High Energy Physics - ICHEP2022\\
  6-13 July, 2022\\
  Bologna, Italy
}

\newcommand{\mr}{\mathrm}
\newcommand{\mc}{\mathcal}

\graphicspath{{figures/}}

\begin{document}
\maketitle

\section{Introduction}
The muon magnetic moment is well predicted in the standard model (SM) of elementary particle physics, and the most accurate calculation is $a_{\mu}^{\mr{SM}}=116591810(43)\times10^{-11}$ \cite{Aoyama:2020ynm}. Its deviation from the SM prediction can be a good probe to new physics. The $(g-2)_{\mu}$ anomaly is first reported by the E821 experiment at BNL \cite{Muong-2:2006rrc}. Last year, the FNAL muon $g-2$ experiment announces the average result $a_{\mu}^{\mr{Exp}}=116592061(41)\times10^{-11}$ after combining the BNL and FNAL data \cite{Muong-2:2021ojo}, which shows the $4.2\sigma$ discrepancy with $\Delta a_{\mu}\equiv a_{\mu}^{\mr{Exp}}-a_{\mu}^{\mr{SM}}=(251\pm59)\times10^{-11}$. There are many interpretations on this anomaly regardless of theoretical and experimental uncertainties, or new physics. In our paper \cite{He:2021yck}, we propose the simultaneous scalar LQ and VLQ extended models to explain this anomaly.
\section{The LQ and VLQ extended models}
For the mediator with mass scale $\Lambda$ above TeV, we have the rough estimation $m_{\mu}^2/(8\pi^2\Lambda^2)\lesssim10^{-10}$. Thus, the chiral enhancements are required to explain the $(g-2)_{\mu}$. In the minimal LQ models, only the $R_2$ and $S_1$ representations can lead to the left and right-handed (non-chiral) couplings to muons at the same time. In fact, the chiral enhancements are induced by the up-type quarks \cite{Dorsner:2016wpm}.

There are seven typical VLQs, while we are interested in the five types with top partner $T$ \cite{Aguilar-Saavedra:2013qpa}, namely, the singlet $T_{L,R}$, doublets $(X,T)_{L,R}/(T,B)_{L,R}$, and triplets $(X,T,B)_{L,R}/(T,B,Y)_{L,R}$. Here, the $X,T,B,Y$ quarks carry the $5/3,2/3,-1/3,-4/3$ electric charges, respectively. Although the five scalar LQs and five $T$ VLQs can result in twenty-five combinations totally, only some combinations can lead to the up-type quark chiral enhancements. In the following, we will study these combinations, which are named as "$\mr{LQ}+\mr{VLQ}$" \footnote{In our paper \cite{He:2021yck}, we also investigate the one LQ and two VLQ extended models.}.

After the electroweak symmetry breaking, there are $t-T$ and $b-B$ mixings with the mixing angles denoted as $\theta_{L,R}^t$ (also $\theta_{L,R}$) and $\theta_{L,R}^b$. Hereafter, the $\sin\theta_{L,R}$ and $\cos\theta_{L,R}$ will be abbreviated as $s_{L,R}$ and $c_{L,R}$ (similar to the $b$). For the mentioned VLQs, there is only one independent mixing angle except for the $(T,B)_{L,R}$ with two independent mixing angles $\theta_R^t$ and $\theta_R^b$. In our paper \cite{He:2021yck}, we list the relevant input parameters and mixing angle identities. For the singlet and triplet VLQs, the $\theta_L$ is chosen as the input mixing angle. For the doublet VLQs, the $\theta_R$ is chosen as the input mixing angle \cite{Aguilar-Saavedra:2013qpa}. In Tab. \ref{tab:LQ+VLQ:LQmut}, we parametrize the couplings in front of the $\bar{\mu}t(R_2^{5/3})^\ast$, $\bar{\mu}T(R_2^{5/3})^\ast$, $\bar{\mu}t^C(S_1)^\ast$, $\bar{\mu}T^C(S_1)^\ast$, $\bar{\mu}t^C(S_3^{1/3})^\ast$, and $\bar{\mu}T^C(S_3^{1/3})^\ast$ interactions.
\begin{table}[!htb]\small
\begin{center}
\begin{tabular}{c|c|c|c|c|c}
\hline
LQ & VLQ & $\overline{\mu_R}t_L$ & $\overline{\mu_L}t_R$ & $\overline{\mu_R}T_L$ & $\overline{\mu_L}T_R$ \\
\hline
\multirow{5}{*}{$R_2$} & $T_{L,R}$ & $y_L^{R_2\mu t}c_L$ & $y_R^{R_2\mu t}c_R-y_R^{R_2\mu T}s_R$ & $y_L^{R_2\mu t}s_L$ & $y_R^{R_2\mu t}s_R+y_R^{R_2\mu T}c_R$ \\
\cline{2-6}
& $(X,T)_{L,R}$ & $y_L^{R_2\mu t}c_L$ & $y_R^{R_2\mu t}c_R$ & $y_L^{R_2\mu t}s_L$ & $y_R^{R_2\mu t}s_R$ \\
\cline{2-6}
& $(T,B)_{L,R}$ & $y_L^{R_2\mu t}c_L-y_L^{R_2\mu T}s_L$ & $y_R^{R_2\mu t}c_R$ & $y_L^{R_2\mu t}s_L+y_L^{R_2\mu T}c_L$ & $y_R^{R_2\mu t}s_R$ \\
\cline{2-6}
& $(X,T,B)_{L,R}$ & $y_L^{R_2\mu t}c_L$ & $y_R^{R_2\mu t}c_R-y_R^{R_2\mu T}s_R$ & $y_L^{R_2\mu t}s_L$ & $y_R^{R_2\mu t}s_R+y_R^{R_2\mu T}c_R$ \\
\cline{2-6}
& $(T,B,Y)_{L,R}$ & $y_L^{R_2\mu t}c_L$ & $y_R^{R_2\mu t}c_R$ & $y_L^{R_2\mu t}s_L$ & $y_R^{R_2\mu t}s_R$ \\
\hline\hline
LQ & VLQ & $\overline{\mu_R}(t_R)^C$ & $\overline{\mu_L}(t_L)^C$ & $\overline{\mu_R}(T_R)^C$ & $\overline{\mu_L}(T_L)^C$\\
\hline
\multirow{5}{*}{$S_1$} & $T_{L,R}$ & $y_L^{S_1\mu t}c_R-y_L^{S_1\mu T}s_R$ & $y_R^{S_1\mu t}c_L$ & $y_L^{S_1\mu t}s_R+y_L^{S_1\mu T}c_R$ & $y_R^{S_1\mu t}s_L$ \\
\cline{2-6}
& $(X,T)_{L,R}$ & $y_L^{S_1\mu t}c_R$ & $y_R^{S_1\mu t}c_L$ & $y_L^{S_1\mu t}s_R$ & $y_R^{S_1\mu t}s_L$ \\
\cline{2-6}
& $(T,B)_{L,R}$ & $y_L^{S_1\mu t}c_R$ & $y_R^{S_1\mu t}c_L-y_R^{S_1\mu T}s_L$ & $y_L^{S_1\mu t}s_R$ & $y_R^{S_1\mu t}s_L+y_R^{S_1\mu T}c_L$ \\
\cline{2-6}
& $(X,T,B)_{L,R}$ & $y_L^{S_1\mu t}c_R$ & $y_R^{S_1\mu t}c_L$ & $y_L^{S_1\mu t}s_R$ & $y_R^{S_1\mu t}s_L$ \\
\cline{2-6}
& $(T,B,Y)_{L,R}$ & $y_L^{S_1\mu t}c_R$ & $y_R^{S_1\mu t}c_L$ & $y_L^{S_1\mu t}s_R$ & $y_R^{S_1\mu t}s_L$ \\
\hline
$S_3$ & $(X,T,B)_{L,R}$ & $-y_L^{S_3\mu T}s_R$ & $y_R^{S_3\mu t}c_L$ & $y_L^{S_3\mu T}c_R$ & $y_R^{S_3\mu t}s_L$ \\ \hline
\end{tabular}
\caption{The $\mr{LQ}\mu t/T$ couplings in the LQ+VLQ models.}\label{tab:LQ+VLQ:LQmut}
\end{center}
\end{table}
In Tab. \ref{tab:LQ+VLQ:LQmub}, we also list the couplings in front of the $\bar{\mu}b(R_2^{2/3})^\ast$, $\bar{\mu}B(R_2^{2/3})^\ast$, $\bar{\mu}b^C(S_3^{4/3})^\ast$, and $\bar{\mu}B^C(S_3^{4/3})^\ast$ interactions.
\begin{table}[!htb]\small
\begin{center}
\begin{tabular}{c|c|c|c|c|c}
\hline
LQ & VLQ & $\overline{\mu_R}b_L$ & $\overline{\mu_L}b_R$ & $\overline{\mu_R}B_L$ & $\overline{\mu_L}B_R$ \\
\hline
\multirow{5}{*}{$R_2$} & $T_{L,R}$ & $y_L^{R_2\mu t}$ & 0 & $\times$ & $\times$ \\
\cline{2-6}
& $(X,T)_{L,R}$ & $y_L^{R_2\mu t}$ & 0 & $\times$ & $\times$ \\
\cline{2-6}
& $(T,B)_{L,R}$ & $y_L^{R_2\mu t}c_L^b-y_L^{R_2\mu T}s_L^b$ & 0 & $y_L^{R_2\mu T}c_L^b+y_L^{R_2\mu t}s_L^b$ & 0 \\
\cline{2-6}
& $(X,T,B)_{L,R}$ & $y_L^{R_2\mu t}c_L^b$ & $-\sqrt{2}y_R^{R_2\mu T}s_R^b$ & $y_L^{R_2\mu t}s_L^b$ & $\sqrt{2}y_R^{R_2\mu T}c_R^b$ \\
\cline{2-6}
& $(T,B,Y)_{L,R}$ & $y_L^{R_2\mu t}c_L^b$ & 0 & $y_L^{R_2\mu t}s_L^b$ & 0 \\
\hline\hline
LQ & VLQ & $\overline{\mu_R}(b_R)^C$ & $\overline{\mu_L}(b_L)^C$ & $\overline{\mu_R}(B_R)^C$ & $\overline{\mu_L}(B_L)^C$\\
\hline
\multirow{5}{*}{$S_1$} & $T_{L,R}$ & 0 & 0 & $\times$ & $\times$ \\
\cline{2-6}
& $(X,T)_{L,R}$ & 0 & 0 & $\times$ & $\times$ \\
\cline{2-6}
& $(T,B)_{L,R}$ & 0 & 0 & 0 & 0 \\
\cline{2-6}
& $(X,T,B)_{L,R}$ & 0 & 0 & 0 & 0 \\
\cline{2-6}
& $(T,B,Y)_{L,R}$ & 0 & 0 & 0 & 0 \\
\hline
$S_3$ & $(X,T,B)_{L,R}$ & $-y_L^{S_3\mu T}s_R^b$ & $\sqrt{2}y_R^{S_3\mu t}c_L^b$ & $y_L^{S_3\mu T}c_R^b$ & $\sqrt{2}y_R^{S_3\mu t}s_L^b$ \\ \hline
\end{tabular}
\caption{The $\mr{LQ}\mu b/B$ couplings in the LQ+VLQ models. The symbol ``$\times$" means no such interactions.}\label{tab:LQ+VLQ:LQmub}
\end{center}
\end{table}
\section{Contributions to the $(g-2)_{\mu}$}
In all of the mentioned models, there are top and $T$ quark contributions with chiral enhancements. In the $R_2+\mr{VLQ}$ models, there are also $b$ and $B$ quark contributions. In the $S_1+\mr{VLQ}$ models, there are no $b$ or $B$ quark contributions. In the $R_2/S_3+(X,T,B)_{L,R}$ models, the $b$ and $B$ quark contributions are also chirally enhanced. For the models with $X$ quark, the $X$ quark only contributes in the $S_3+(X,T,B)_{L,R}$ model but without the chiral enhancements. For the $R_2/S_1+(T,B,Y)_{L,R}$ models, the $Y$ quark does not contribute to $(g-2)_{\mu}$.

The complete contributions can be obtained from our paper \cite{He:2021yck}. Of course, they are dominated by the chirally enhanced contributions, because the non-chirally enhanced contributions are suppressed by the factor $m_{\mu}/m_t(m_T)\le10^{-3}$. Considering $m_b\ll m_t\ll m_T\approx m_B$, and $s_{L,R}\ll1$, we show the approximate formulae of $\Delta a_{\mu}$ in Tab. \ref{tab:LQ+VLQ:g-2}. In the $R_2+(X,T)_{L,R}/(T,B,Y)_{L,R}$ and $S_1+(X,T)_{L,R}/(X,T,B)_{L,R}/(T,B,Y)_{L,R}$ models, the $T$ contributions are highly suppressed by the factor $m_ts_{L,R}^2/m_T$. In the $R_2+T_{L,R}/(T,B)_{L,R}/(X,T,B)_{L,R}$ and $S_1+T_{L,R}/(T,B)_{L,R}$ models, the $T$ contributions are suppressed by the mixing angle $s_{L,R}$. In the $S_3+(X,T,B)_{L,R}$ model, the $T$ and $B$ quark contributions are dominated than top by the factor $m_T/m_t$.
\begin{table}[!htb]\footnotesize
\begin{center}
\begin{tabular}{>{\scriptsize}p{0.25cm}<{\centering}|>{\scriptsize}p{1.5cm}<{\centering}|c|c}
\hline
LQ & VLQ & the approximate expressions of $\Delta\overline{a_{\mu}}$ & \makecell{coupling product order \\ of $T$ compared to $t$}\\
\hline
\multirow{5}{*}{$R_2$} & $T_{L,R}$ & $\frac{m_T}{m_t}f_{LR}^{R_2}(m_T^2/m_{R_2}^2)\mr{Re}[y_R^{R_2\mu T}(y_L^{R_2\mu t})^\ast]s_L+(\frac{1}{4}+\log \frac{m_t^2}{m_{R_2}^2})\mr{Re}[y_R^{R_2\mu t}(y_L^{R_2\mu t})^\ast]$ & $s_L$ \\
\cline{2-4}
& $(X,T)_{L,R}$ & $\frac{m_T}{m_t}f_{LR}^{R_2}(m_T^2/m_{R_2}^2)\mr{Re}[y_R^{R_2\mu t}(y_L^{R_2\mu t})^\ast]s_Ls_R+(\frac{1}{4}+\log \frac{m_t^2}{m_{R_2}^2})\mr{Re}[y_R^{R_2\mu t}(y_L^{R_2\mu t})^\ast]$ & $m_ts_R^2/m_T$\\
\cline{2-4}
& $(T,B)_{L,R}$ & $\frac{m_T}{m_t}f_{LR}^{R_2}(m_T^2/m_{R_2}^2)\mr{Re}[y_R^{R_2\mu t}(y_L^{R_2\mu T})^\ast]s_R+(\frac{1}{4}+\log \frac{m_t^2}{m_{R_2}^2})\mr{Re}[y_R^{R_2\mu t}(y_L^{R_2\mu t})^\ast]$ & $s_R$\\
\cline{2-4}
\rule{0pt}{20pt}& $(X,T,B)_{L,R}$ & \makecell{$\frac{m_T}{m_t}[f_{LR}^{R_2}(m_T^2/m_{R_2}^2)+2\widetilde{f}_{LR}^{R_2}(m_T^2/m_{R_2}^2)]\cdot\mr{Re}[y_R^{R_2\mu T}(y_L^{R_2\mu t})^\ast]s_L$\\$+(\frac{1}{4}+\log \frac{m_t^2}{m_{R_2}^2})\mr{Re}[y_R^{R_2\mu t}(y_L^{R_2\mu t})^\ast]$} & $s_L$\\
\cline{2-4}
& $(T,B,Y)_{L,R}$ & $\frac{m_T}{m_t}f_{LR}^{R_2}(m_T^2/m_{R_2}^2)\mr{Re}[y_R^{R_2\mu t}(y_L^{R_2\mu t})^\ast]s_Ls_R+(\frac{1}{4}+\log \frac{m_t^2}{m_{R_2}^2})\mr{Re}[y_R^{R_2\mu t}(y_L^{R_2\mu t})^\ast]$ & $m_ts_L^2/m_T$\\
\hline\hline
\multirow{5}{*}{$S_1$} & $T_{L,R}$ & $\frac{m_T}{m_t}f_{LR}^{S_1}(m_T^2/m_{S_1}^2)\mr{Re}[y_L^{S_1\mu T}(y_R^{S_1\mu t})^\ast]s_L-(\frac{7}{4}+\log \frac{m_t^2}{m_{S_1}^2})\mr{Re}[y_L^{S_1\mu t}(y_R^{S_1\mu t})^\ast]$ & $s_L$\\
\cline{2-4}
& $(X,T)_{L,R}$ & $\frac{m_T}{m_t}f_{LR}^{S_1}(m_T^2/m_{S_1}^2)\mr{Re}[y_L^{S_1\mu t}(y_R^{S_1\mu t})^\ast]s_Ls_R-(\frac{7}{4}+\log \frac{m_t^2}{m_{S_1}^2})\mr{Re}[y_L^{S_1\mu t}(y_R^{S_1\mu t})^\ast]$ & $m_ts_R^2/m_T$\\
\cline{2-4}
& $(T,B)_{L,R}$ & $\frac{m_T}{m_t}f_{LR}^{S_1}(m_T^2/m_{S_1}^2)\mr{Re}[y_L^{S_1\mu t}(y_R^{S_1\mu T})^\ast]s_R-(\frac{7}{4}+\log \frac{m_t^2}{m_{S_1}^2})\mr{Re}[y_L^{S_1\mu t}(y_R^{S_1\mu t})^\ast]$ & $s_R$\\
\cline{2-4}
& $(X,T,B)_{L,R}$ & $\frac{m_T}{m_t}f_{LR}^{S_1}(m_T^2/m_{S_1}^2)\mr{Re}[y_L^{S_1\mu t}(y_R^{S_1\mu t})^\ast]s_Ls_R-(\frac{7}{4}+\log \frac{m_t^2}{m_{S_1}^2})\mr{Re}[y_L^{S_1\mu t}(y_R^{S_1\mu t})^\ast]$ & $m_ts_L^2/m_T$\\
\cline{2-4}
& $(T,B,Y)_{L,R}$ & $\frac{m_T}{m_t}f_{LR}^{S_1}(m_T^2/m_{S_1}^2)\mr{Re}[y_L^{S_1\mu t}(y_R^{S_1\mu t})^\ast]s_Ls_R-(\frac{7}{4}+\log \frac{m_t^2}{m_{S_1}^2})\mr{Re}[y_L^{S_1\mu t}(y_R^{S_1\mu t})^\ast]$ & $m_ts_L^2/m_T$\\
\hline
\rule{0pt}{20pt}$S_3$ & $(X,T,B)_{L,R}$ & \makecell{$\frac{m_T}{m_t}[f_{LR}^{S_3}(m_T^2/m_{S_3}^2)+2\widetilde{f}_{LR}^{S_3}(m_T^2/m_{S_3}^2)]\cdot\mr{Re}[y_L^{S_3\mu T}(y_R^{S_3\mu t})^\ast]s_L$\\$+(\frac{7}{4}+\log \frac{m_t^2}{m_{S_3}^2})\mr{Re}[y_L^{S_3\mu T}(y_R^{S_3\mu t})^\ast]s_R$} & $m_T/m_t$\\ \hline
\end{tabular}
\caption{In the third column, we show the approximate formulae of the $\Delta a_{\mu}$. In the fourth column, we show the order of the multiplication of left and right-handed $T$ LQ Yukawa couplings with respect to the top quark. In the above, we redefine $\Delta a_{\mu}$ as $m_{\mu}m_t\Delta\overline{a_{\mu}}/(4\pi^2m_{\mr{LQ}}^2)$.} \label{tab:LQ+VLQ:g-2}
\end{center}
\end{table}
\section{Numerical analysis}
\begin{figure}[!htb]
\begin{center}
\includegraphics[scale=0.2]{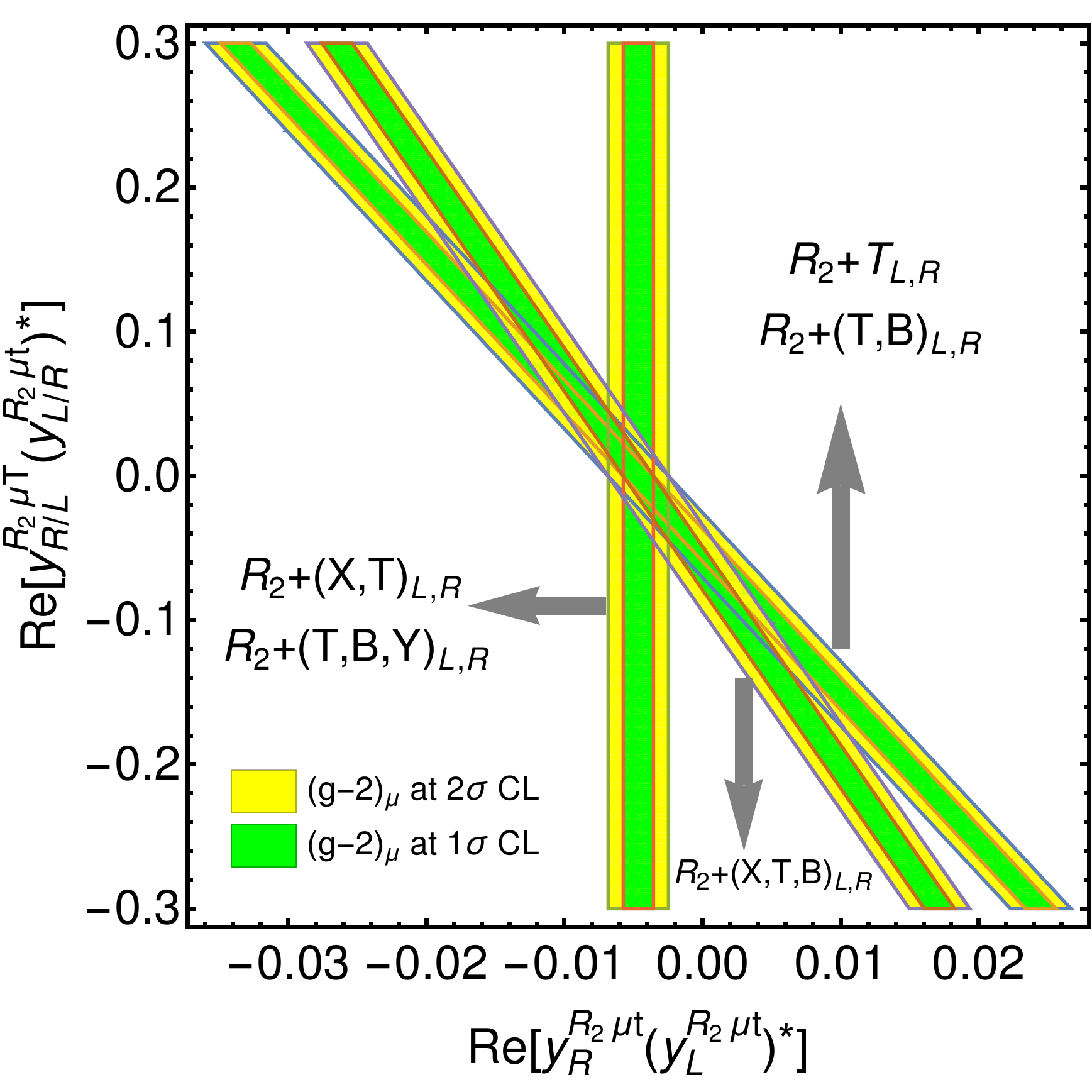}\qquad
\includegraphics[scale=0.2]{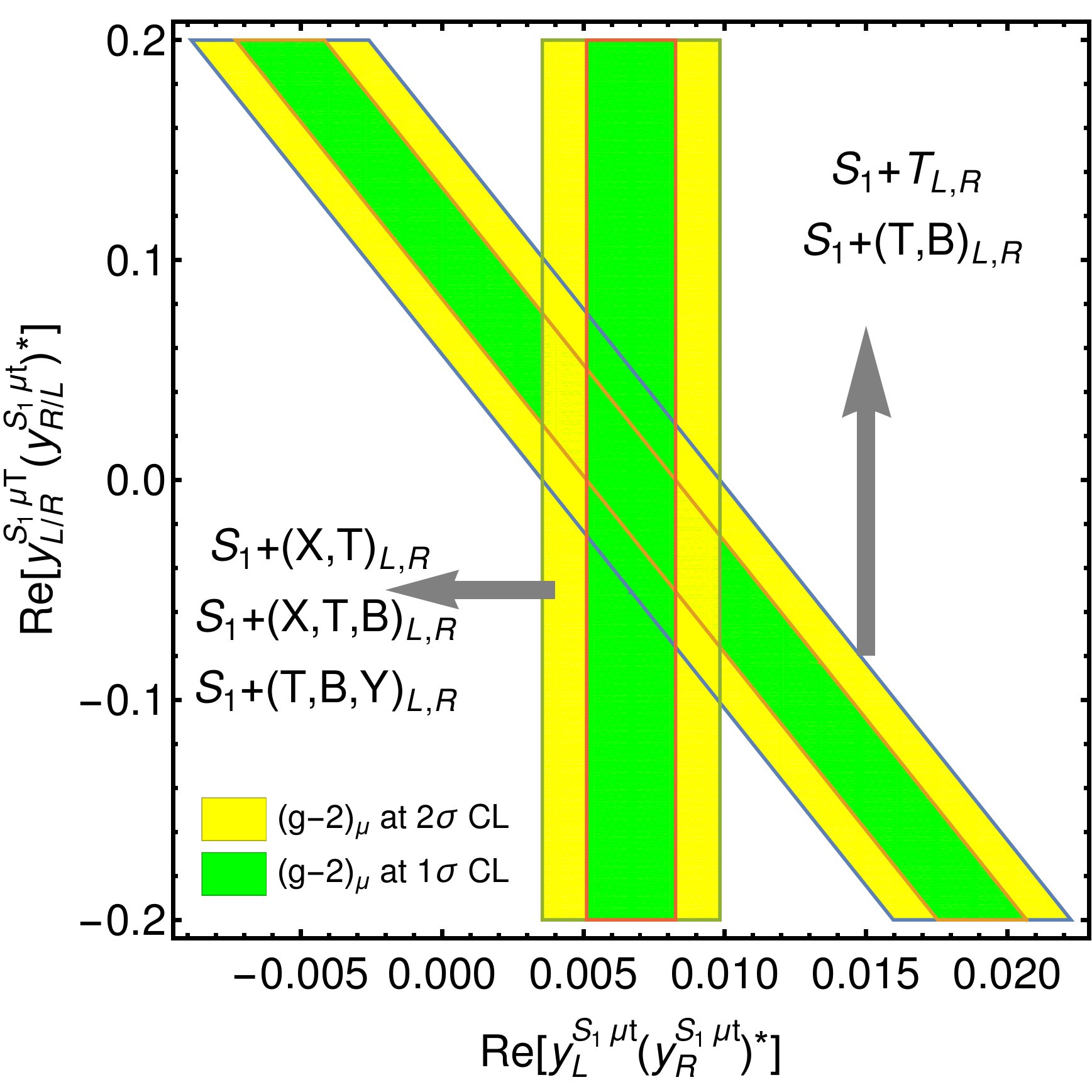}\qquad
\includegraphics[scale=0.2]{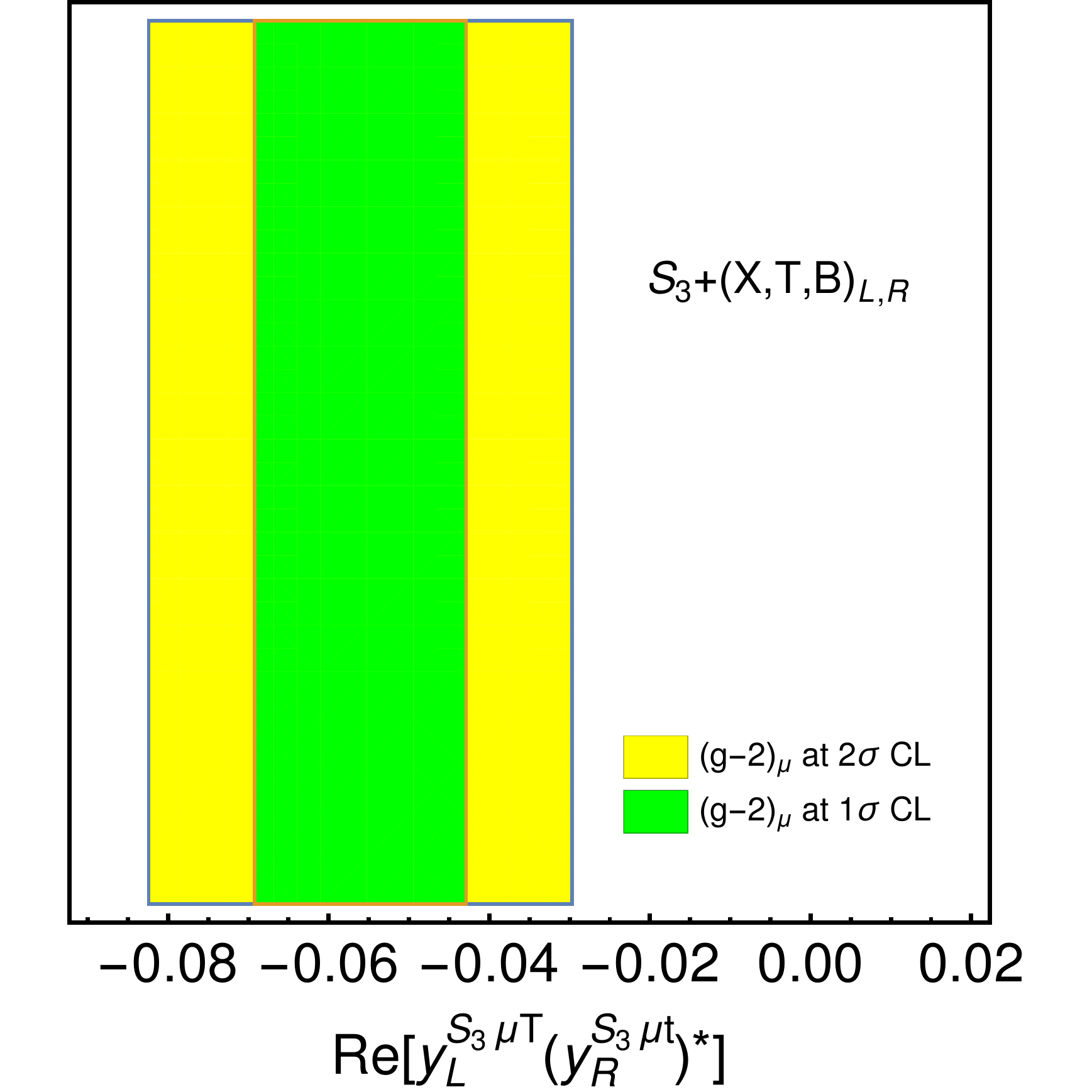}
\caption{\footnotesize The allowed region in the plane of $\mr{Re}[y_{R/L}^{R_2\mu T}(y_{L/R}^{R_2\mu t})^\ast]-\mr{Re}[y_R^{R_2\mu t}(y_L^{R_2\mu t})^\ast]$ (left, $R_2+\mr{VLQ}$) and $\mr{Re}[y_{L/R}^{S_1\mu T}(y_{R/L}^{S_1\mu t})^\ast]-\mr{Re}[y_L^{S_1\mu t}(y_R^{S_1\mu t})^\ast]$ (middle, $S_1+\mr{VLQ}$). The right is for the $S_3+(X,T,B)_{L,R}$ model.}\label{fig:LQ+VLQ:g-2num}
\end{center}
\end{figure}

We choose the input parameters as $m_\mu=105.66\mathrm{MeV}$, $m_b=4.2\mr{GeV}$, and $m_t=172.5\mr{GeV}$ \cite{ParticleDataGroup:2020ssz}. For the VLQ parameters, the main constraints are from direct search \cite{CMS:2018wpl, ATLAS:2018ziw} and electro-weak precision observables \cite{Aguilar-Saavedra:2013qpa, Chen:2017hak}, which require the VLQ mass to be $\mc{O}(\mr{TeV})$ and the input mixing angle to be less than 0.1. For the LQ mass, the direct search requires it to be above $\mr{TeV}$ \cite{CMS:2018oaj, ATLAS:2020xov}. Then, we adopt the mass parameters to be $m_T=1\mr{TeV}$ and $m_{\mr{LQ}}=2\mr{TeV}$ by default. The input mixing angle is set as $s_L=0.05$ (singlet and triplet VLQs) and $s_R=0.05$ (doublet VLQs). In Fig. \ref{fig:LQ+VLQ:g-2num}, we show the regions allowed at $1\sigma$ (green) and $2\sigma$ (yellow) CL, respectively.

\section{LQ Phenomenology at hadron colliders}
In the minimal $R_2/S_1$ models, the LQ decay final states are SM quark and lepton. In the LQ+VLQ models, there are new LQ decay channels. Here, we will study the $R_2^{5/3}\rightarrow t/T\mu^+$ and $S_1/S_3^{1/3}\rightarrow\bar{t}/\bar{T}\mu^+$ decay channels \footnote{For the other LQs, the decay channels can be $R_2^{2/3}\rightarrow b/B\mu^+$, $S_3^{4/3}\rightarrow \bar{b}/\bar{B}\mu^+$, and $S_3^{-2/3}\rightarrow \bar{X}\mu^+$.}. Considering $m_t\ll m_T$ and $s_{L,R}\ll1$, we show the approximate expressions of $\Gamma(\mr{LQ}\rightarrow T\mu)/\Gamma(\mr{LQ}\rightarrow t\mu)$ in Tab. \ref{tab:LQ+VLQ:widthapp}. Then, we find that the $T\mu$ decay channel is important in the $R_2+T_{L,R}/(T,B)_{L,R}/(X,T,B)_{L,R}$, $S_1+T_{L,R}/(T,B)_{L,R}$, and $S_3+(X,T,B)_{L,R}$ models.
\begin{table}[!htb]
\begin{center}\begin{scriptsize}
\begin{tabular}{c|c|c|c}
\hline
LQ & VLQ & the approximate expressions of $\Gamma(\mr{LQ}\rightarrow T\mu)/\Gamma(\mr{LQ}\rightarrow t\mu)$ & suppress or not\\
\hline
\multirow{5}{*}{$R_2$} & $T_{L,R}$ & $(1-\frac{m_T^2}{m_{R_2}^2})^2|y_R^{R_2\mu T}|^2/(|y_L^{R_2\mu t}|^2+|y_R^{R_2\mu t}|^2)$ & No\\
\cline{2-4}
& $(X,T)_{L,R}$ & $(1-\frac{m_T^2}{m_{R_2}^2})^2|y_R^{R_2\mu t}|^2s_R^2/(|y_L^{R_2\mu t}|^2+|y_R^{R_2\mu t}|^2)$ & $s_R^2$\\
\cline{2-4}
& $(T,B)_{L,R}$ & $(1-\frac{m_T^2}{m_{R_2}^2})^2|y_L^{R_2\mu T}|^2/(|y_L^{R_2\mu t}|^2+|y_R^{R_2\mu t}|^2)$ & No\\
\cline{2-4}
& $(X,T,B)_{L,R}$ & $(1-\frac{m_T^2}{m_{R_2}^2})^2|y_R^{R_2\mu T}|^2/(|y_L^{R_2\mu t}|^2+|y_R^{R_2\mu t}|^2)$ & No\\
\cline{2-4}
& $(T,B,Y)_{L,R}$ & $(1-\frac{m_T^2}{m_{R_2}^2})^2|y_L^{R_2\mu t}|^2s_L^2/(|y_L^{R_2\mu t}|^2+|y_R^{R_2\mu t}|^2)$ & $s_L^2$\\
\hline\hline
\multirow{5}{*}{$S_1$} & $T_{L,R}$ & $(1-\frac{m_T^2}{m_{S_1}^2})^2|y_L^{S_1\mu T}|^2/(|y_L^{S_1\mu t}|^2+|y_R^{S_1\mu t}|^2)$ & No\\
\cline{2-4}
& $(X,T)_{L,R}$ & $(1-\frac{m_T^2}{m_{S_1}^2})^2|y_L^{S_1\mu t}|^2s_R^2/(|y_L^{S_1\mu t}|^2+|y_R^{S_1\mu t}|^2)$ & $s_R^2$\\
\cline{2-4}
& $(T,B)_{L,R}$ & $(1-\frac{m_T^2}{m_{S_1}^2})^2|y_R^{S_1\mu T}|^2/(|y_L^{S_1\mu t}|^2+|y_R^{S_1\mu t}|^2)$ & No\\
\cline{2-4}
& $(X,T,B)_{L,R}$ & $(1-\frac{m_T^2}{m_{S_1}^2})^2|y_R^{S_1\mu t}|^2s_L^2/(|y_L^{S_1\mu t}|^2+|y_R^{S_1\mu t}|^2)$ & $s_L^2$\\
\cline{2-4}
& $(T,B,Y)_{L,R}$ & $(1-\frac{m_T^2}{m_{S_1}^2})^2|y_R^{S_1\mu t}|^2s_L^2/(|y_L^{S_1\mu t}|^2+|y_R^{S_1\mu t}|^2)$ & $s_L^2$\\
\hline
$S_3$ & $(X,T,B)_{L,R}$ & $(1-\frac{m_T^2}{m_{S_3}^2})^2|y_L^{S_3\mu T}|^2/|y_R^{S_3\mu t}|^2$ & No\\ \hline
\end{tabular}
\caption{\footnotesize In the third column, we list the approximate formulae of $T\mu$ partial decay width over $t\mu$ in the LQ+VLQ models. In the fourth column, we show the order of $\Gamma(\mr{LQ}\rightarrow T\mu)$ compared to the $\Gamma(\mr{LQ}\rightarrow t\mu)$.} \label{tab:LQ+VLQ:widthapp}
\end{scriptsize}\end{center}
\end{table}
For the LQ production, there are pair, single, and off-shell channels. What is more, the $T$ quark can decay into the $bW,tZ,th$ final states further. Thus, it will lead to the characteristic multi-top and multi-muon signals at hadron colliders.
\section{Summary and conclusions}
We explain the $(g-2)_{\mu}$ anomaly in the LQ and VLQ extended models. In the $R_2+(X,T)_{L,R}/(T,B,Y)_{L,R}$ and $S_1+(X,T)_{L,R}/(X,T,B)_{L,R}/(T,B,Y)_{L,R}$ models, it is dominated by the top quark contributions. In the $R_2+T_{L,R}/(T,B)_{L,R}/(X,T,B)_{L,R}$ and $S_1+T_{L,R}/(T,B)_{L,R}$ models, both the top and $T$ quark contributions are important. In the $S_3+(X,T,B)_{L,R}$ model, it is dominated by the $T$ and $B$ quark contributions. In addition to the conventional $t\mu$ decay channel, the LQ can also decay into $T\mu$ final states, which can become important in the $R_2+T_{L,R}/(T,B)_{L,R}/(X,T,B)_{L,R}$, $S_1+T_{L,R}/(T,B)_{L,R}$ models, and $S_3+(X,T,B)_{L,R}$ models.

\begin{acknowledgments}\small
This research was supported by an appointment to the Young Scientist Training Program at the APCTP through the Science and Technology Promotion Fund and Lottery Fund of the Korean Government. This was also supported by the Korean Local Governments-Gyeongsangbuk-do Province and Pohang City.
\end{acknowledgments}

%
\bibliographystyle{unsrt}
\bibliography{PoS_ICHEP2022_137_Shi-PingHe}

\begin{thebibliography}{10}

\bibitem{Aoyama:2020ynm}
T.~Aoyama et~al.
\newblock {The anomalous magnetic moment of the muon in the Standard Model}.
\newblock {\em Phys. Rept.}, 887:1--166, 2020.

\bibitem{Muong-2:2006rrc}
G.~W. Bennett et~al.
\newblock {Final Report of the Muon E821 Anomalous Magnetic Moment Measurement
  at BNL}.
\newblock {\em Phys. Rev. D}, 73:072003, 2006.

\bibitem{Muong-2:2021ojo}
B.~Abi et~al.
\newblock {Measurement of the Positive Muon Anomalous Magnetic Moment to 0.46
  ppm}.
\newblock {\em Phys. Rev. Lett.}, 126(14):141801, 2021.

\bibitem{He:2021yck}
Shi-Ping He.
\newblock {Leptoquark and vectorlike quark extended models as the explanation
  of the muon $g-2$ anomaly}.
\newblock {\em Phys. Rev. D}, 105(3):035017, 2022.
\newblock [Erratum: Phys.Rev.D 106, 039901 (2022)].

\bibitem{Dorsner:2016wpm}
I.~Dor\v{s}ner, S.~Fajfer, A.~Greljo, J.~F. Kamenik, and N.~Ko\v{s}nik.
\newblock {Physics of leptoquarks in precision experiments and at particle
  colliders}.
\newblock {\em Phys. Rept.}, 641:1--68, 2016.

\bibitem{Aguilar-Saavedra:2013qpa}
J.~A. Aguilar-Saavedra, R.~Benbrik, S.~Heinemeyer, and M.~P\'erez-Victoria.
\newblock {Handbook of vectorlike quarks: Mixing and single production}.
\newblock {\em Phys. Rev. D}, 88(9):094010, 2013.

\bibitem{ParticleDataGroup:2020ssz}
P.~A. Zyla et~al.
\newblock {Review of Particle Physics}.
\newblock {\em PTEP}, 2020(8):083C01, 2020.

\bibitem{CMS:2018wpl}
Albert~M Sirunyan et~al.
\newblock {Search for vector-like quarks in events with two oppositely charged
  leptons and jets in proton-proton collisions at $\sqrt{s} =$ 13 TeV}.
\newblock {\em Eur. Phys. J. C}, 79(4):364, 2019.

\bibitem{ATLAS:2018ziw}
Morad Aaboud et~al.
\newblock {Combination of the searches for pair-produced vector-like partners
  of the third-generation quarks at $\sqrt{s} =$ 13 TeV with the ATLAS
  detector}.
\newblock {\em Phys. Rev. Lett.}, 121(21):211801, 2018.

\bibitem{Chen:2017hak}
Chien-Yi Chen, S.~Dawson, and Elisabetta Furlan.
\newblock {Vectorlike fermions and Higgs effective field theory revisited}.
\newblock {\em Phys. Rev. D}, 96(1):015006, 2017.

\bibitem{CMS:2018oaj}
Albert~M Sirunyan et~al.
\newblock {Search for leptoquarks coupled to third-generation quarks in
  proton-proton collisions at $\sqrt{s}=$ 13 TeV}.
\newblock {\em Phys. Rev. Lett.}, 121(24):241802, 2018.

\bibitem{ATLAS:2020xov}
Georges Aad et~al.
\newblock {Search for pair production of scalar leptoquarks decaying into
  first- or second-generation leptons and top quarks in
  proton\textendash{}proton collisions at $\sqrt{s}$ = 13 TeV with the ATLAS
  detector}.
\newblock {\em Eur. Phys. J. C}, 81(4):313, 2021.

\end{thebibliography}

\end{document}